\documentclass[aps,prb,twocolumn,superscriptaddress,showpacs,floatfix]{revtex4-2}
\usepackage{amsmath,amsfonts,amssymb}
\usepackage{graphicx}
\usepackage[colorlinks=true,linkcolor=blue,citecolor=blue,urlcolor=blue]{hyperref}
\usepackage{booktabs}
\usepackage{siunitx}
\usepackage{physics}
\usepackage{subcaption}
\usepackage{multirow}
\usepackage{threeparttable}
\usepackage[makeroom]{cancel}
\usepackage{ragged2e}
\usepackage{bbm}
\usepackage{float}
\usepackage{amsmath}
\usepackage{lipsum}

\begin{document}

\title{Active learning molecular beam epitaxy of complex quantum materials}

\author{Raghutheja Bollampally}
\thanks{Equal contribution}
\affiliation{Department of Physics, The Hong Kong University of Science and Technology, Clear Water Bay, Kowloon, Hong Kong}
\author{Soumya Sankar}
\thanks{Equal contribution}
\affiliation{Department of Physics, The Hong Kong University of Science and Technology, Clear Water Bay, Kowloon, Hong Kong}
\author{Yuqi Qin}
\thanks{Equal contribution}
\affiliation{Department of Physics, The Hong Kong University of Science and Technology, Clear Water Bay, Kowloon, Hong Kong}
\author{Berthold J\"{a}ck}
\email{bjaeck@ust.hk}
\affiliation{Department of Physics, The Hong Kong University of Science and Technology, Clear Water Bay, Kowloon, Hong Kong}

\date{\today}

\begin{abstract}
The integration of machine learning (ML) into materials science offers a transformative pathway toward fully autonomous synthesis workflows. For precise thin-film deposition techniques like molecular beam epitaxy (MBE), this automation is critical to overcome the time-consuming, manual navigation of high-dimensional thermodynamic phase spaces. Existing approaches for ML-assisted thin film growth predominantly rely on continuous Bayesian optimization (BO) models that assume smooth parameter landscapes. Consequently, they struggle to capture the abrupt crystallographic phase boundaries and narrow growth windows inherent to binary quantum materials. Here, we demonstrate an active learning protocol based on Sequential Model-Based Optimization (SMBO) designed specifically for the closed-loop MBE of such compounds. To overcome the limitations of continuous models while retaining the efficient exploration-exploitation logic of traditional BO, we combine a random forest surrogate model capable of capturing highly non-linear phase transitions and thermodynamics constraints of the growth process with an expected improvement function to predict optimum growth parameters. We apply this combined SMBO framework to the MBE of the topological Weyl ferromagnet Fe$_3$Sn, which exists as a metastable line compound. Using a small initial training set of fewer than twenty growth iterations, our active learning loop rapidly navigates a complex optimization landscape to identify an optimum growth window bounded by sharp transitions. Within only four active learning iterations, the absolute predictive error is halved to $\approx10\%$. This data-efficient framework paves the way for the autonomous discovery and thin-film synthesis of functional quantum materials.
\end{abstract}

\maketitle

\section{Introduction}

Generative artificial intelligence (AI) models and machine learning (ML) techniques offer unprecedented opportunities for autonomous materials discovery and the optimization of underlying synthesis workflows~\cite{osada2020adaptive,10.1039/d4mh01909a,moses26,messecar2024quantum,zaki2025self}. AI-driven material discovery~\cite{merchant2023scaling, cai2020machine} has the potential to dramatically accelerate the synthesis of novel functional materials, such as multiferroics, complex magnetic compounds, and topological materials. Recent advances in autonomous synthesis and in-situ diagnostics have already succeeded in the autonomous solid-state synthesis of inorganic powders using computations, historical data, and active learning to plan and interpret experiments performed via robotics~\cite{szymanski2023autonomous}. Similarly, autonomous experimentation platforms have been developed for colloidal nano-particle synthesis and the fabrication of 2D material devices~\cite{kusne2020fly, shi2026qumus}.

In addition to these high-level generative AI approaches, which drive the field toward autonomous discovery protocols~\cite{kutsukake2024review}, data-driven workflow optimization is of particular interest for physical vapor deposition techniques like molecular beam epitaxy (MBE)~\cite{shen2026demand,shrivastava2024bayesian}. Unlike most bulk synthesis methods, the MBE of thin films occurs in thermal non-equilibrium and is governed by a delicate set of control parameters, such as substrate temperature and evaporant stoichiometry, that determine the surface kinetics and chemistry of the film growth process~\cite{herman2012molecular}. Consequently, realizing and optimizing the MBE of complex functional materials requires navigating a highly sensitive, multi-dimensional phase space.

\begin{figure}[t] % Notice the asterisk (*) here and the [t] specifier
\centering
\includegraphics[width=8.5cm]{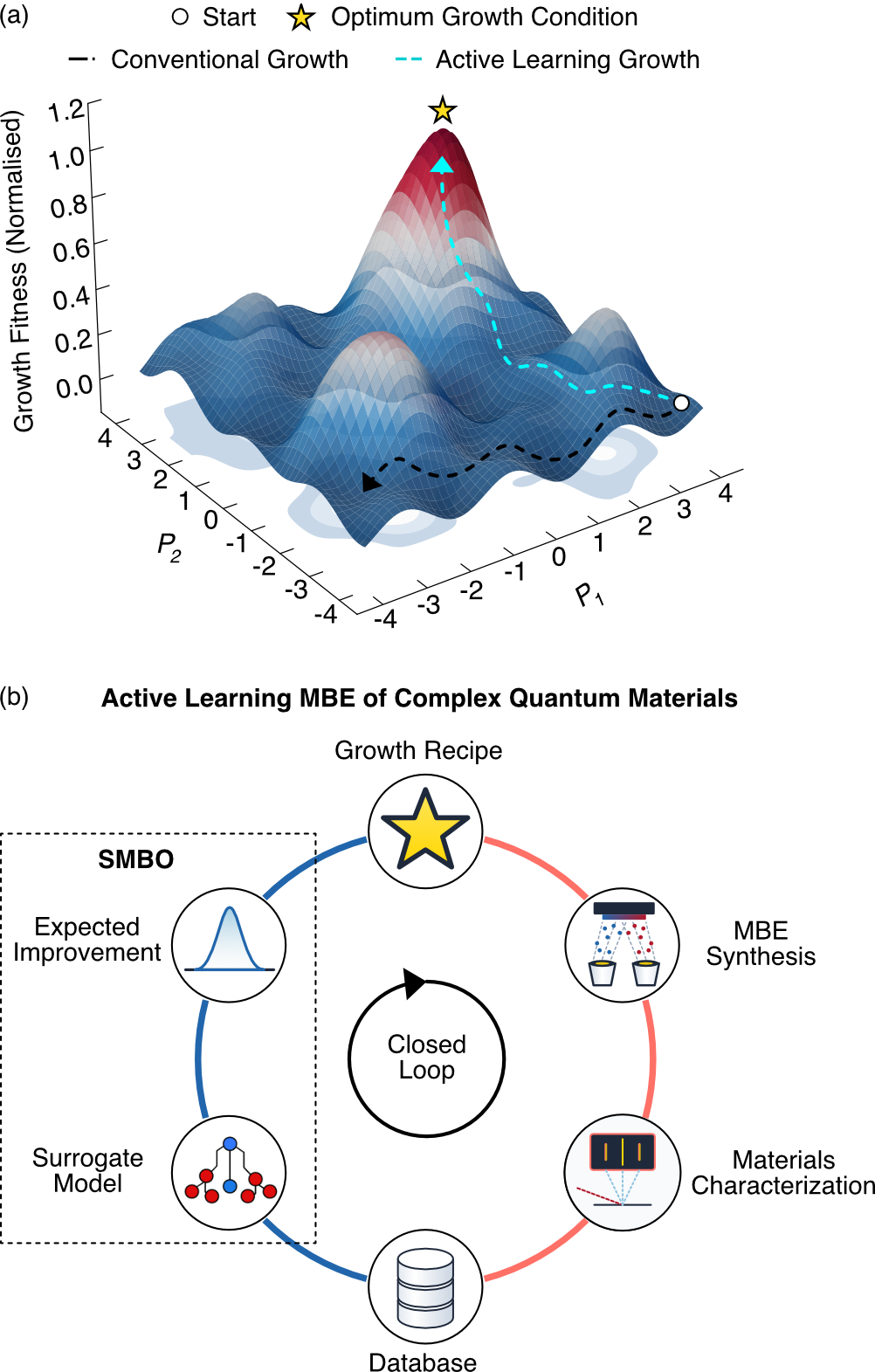}
    \caption{\justifying\textbf{Navigating complex thin-film growth landscapes via closed-loop active learning using an sequential model based optimization.} (a) Schematic representation of the normalized growth fitness landscape over synthesis parameters $P_1$ and $P_2$. The optimization surface of thin film growth is characterized by non-convex curvature with local maxima and minima. Shown are schematic optimization trajectories of conventional (red dashed line) and active learning (turquoise dashed line) growth. (b) Schematic representation of active learning MBE using a sequential model-based optimization (SMBO) framework. The closed loop cycles through physical deposition, structural characterization, and database expansion. These empirical data continuously train a predictive surrogate model, which is then evaluated by an expected improvement function to propose the optimal growth recipe for the next experimental iteration.}
    \label{fig:fig1}
\end{figure}

Conventionally, this phase space is navigated manually by varying one control parameter $P_i$ at a time while keeping others fixed [see Fig.\ref{fig:fig1}(a)]. This approach typically requires dozens of iterations and relies heavily on the intuition of individual researchers. To overcome this critical bottleneck and reach the optimum growth condition faster, ML techniques are increasingly being adopted. For example, Bayesian optimization (BO)~\cite{brochu2010tutorial, shahriari2015taking} has successfully optimized the MBE of several materials by analyzing crystallographic and electrical properties~\cite{wakabayashi2019machine, ohkubo2021realization, wakabayashi2022bayesian, wakabayashi2023stoichiometric, trice2026machine}. More recently, unsupervised learning and convolutional neural networks applied to reflective high-energy electron diffraction (RHEED) patterns have demonstrated the ability to predict film crystallinity and growth modes~\cite{yu2025multimodal, muetzel2026rheed}.

These results collectively showcase that ML techniques can model the MBE of thin film materials through a data-driven analysis of spectroscopic data. While these algorithmic advances successfully reduce growth iterations, integrating these ML techniques into an end-to-end active learning pipeline—where predictive surrogate models dynamically interact with experimental execution to completely remove human intuition from the loop—remains a critical challenge. Existing proof-of-principle demonstrations based on BO have been restricted to structurally simple, single-element thin films (e.g., Ag or TiN)\cite{ohkubo2021realization, zheng2025self} with typically low-dimensional optimization landscapes. Consequently, it remains an open question whether active learning can successfully navigate the high-dimensional yet physically constrained thermodynamic phase space required to synthesize complex binary compounds, where narrow growth windows and competing secondary phases make manual optimization exceedingly difficult. Furthermore, standard BO approaches typically assume smooth, continuous parameter spaces and often struggle to capture the abrupt crystallographic phase boundaries and narrow growth windows that separate competing secondary phases in these materials\cite{garnett2023bayesian}.

In this study, we address this critical gap by demonstrating an active learning (AL) protocol based on Sequential Model-Based Optimization (SMBO)~\cite{hutter2011sequential} for the closed-loop MBE of the topological Weyl ferromagnet Fe$_3$Sn~\cite{sankar2025room}, which exists as a metastable line compound within a complex equilibrium Fe-Sn phase diagram~\cite{shen2022thermodynamical}. To navigate the complex optimization landscape of the MBE of this material, our SMBO framework leverages the efficient exploration-exploitation logic of traditional BO via an expected improvement (EI) acquisition function. Crucially, to overcome the limitations of standard continuous models, we replace the conventional Gaussian process with a random forest (RF) surrogate model~\cite{hastie2009elements}. Specifically chosen for its ability to naturally capture non-linearities and sharp phase boundaries, the RF surrogate is trained on an initial data set to predict optimal growth parameters for subsequent thin film samples. The stoichiometry and crystallographic properties of these samples are then analyzed, and the results are fed back to retrain the model, closing the active learning loop~[see Fig.\ref{fig:fig1}(b)]. We show that an initial training on fewer than twenty growth iterations, guided by this combined RF-EI framework, successfully identifies the optimum growth window and gradually reduces the absolute predictive error by a factor of two to $\approx10\%$ within only four additional growth iterations. Our results demonstrate the AL of MBE for complex quantum materials and pave the way toward the fully autonomous MBE of functional thin film materials using automated synthesis and analysis tools.

 % Notice the asterisk (*) here too

This paper is organized as follows. Section II provides an overview of the experimental methods, including MBE synthesis, characterization techniques, the feature selection and data processing, and the SMBO active learning framework. Section III presents the experimental results, including model performance metrics, convergence behavior, the identification of optimal growth conditions, and the demonstration of active learning MBE of complex quantum materials. Section IV discusses the implications of our findings and concludes with a summary and outlook for future work.

\section{Methods}

\subsection{Molecular Beam Epitaxy and Experimental Characterization of Fe$_3$Sn}
The intermetallic kagome metal Fe$_3$Sn is characterized by topological Weyl points in its electronic structure~\cite{belbase2023large}, giving rise to a room-temperature anomalous in-plane Hall effect~\cite{sankar2025room} that could facilitate advanced magnetic sensing~\cite{zheng2026dominant}. Crucially for this study, the solid solution of Fe$_3$Sn exists as a metastable line compound over a narrow temperature range ($\approx 100$\,K) within the Fe-Sn equilibrium phase diagram~\cite{shen2022thermodynamical}. Its phase stability is tightly bounded by competing homologous compounds (Fe$_x$Sn$_y$) as a function of stoichiometry. This complex thermodynamic landscape results in an exceptionally narrow growth window, making the manual optimization of single-phase, high-quality Fe$_3$Sn films exceedingly difficult~\cite{chatterjee2025challenges}. This sensitivity to multi-dimensional parameter variations renders Fe$_3$Sn an ideal testbed for validating our closed-loop AL-MBE framework on complex quantum materials.

Epitaxial Fe$_3$Sn films were grown in a custom MBE system on $c$-plane sapphire substrates coated with a 5\,nm thick Pt(111) buffer layer~\cite{cheng2022atomic, sankar2025room}. Following standard chemical cleaning and thermal annealing (1300\,K in air) to achieve atomically flat step-terraces, the substrates were degassed in ultra-high vacuum ($p \leq 5 \times 10^{-10}$\,mbar) prior to Pt deposition. For the Fe$_3$Sn active layer, high-purity Fe and Sn were co-evaporated from effusion cells at a nominal rate of 1.2\,\AA/min. To define the phase space for the AL optimization, the Fe-to-Sn beam flux ratio ($J_{\text{Fe}}/J_{\text{Sn}}$) was varied between 0.5 and 1.2, while the substrate heater power was varied between 3.0 and 4.7\,W (corresponding to substrate temperatures of roughly 100--215\,$^\circ$C). The growth process was monitored \textit{in-situ} via RHEED, followed by \textit{ex-situ} X-ray diffraction (XRD) and energy-dispersive X-ray spectroscopy (EDS) to quantify bulk crystallinity and chemical stoichiometry, respectively.

To initialize the active learning workflow, a sparse dataset of 17 historical growth iterations was compiled. These initial films were grown using conventional, human-intuition-based parameter sweeps to map the broad boundaries of the phase space based on RHEED, XRD, and EDS feedback. Subsequently, four additional autonomous growth iterations were executed, entirely guided by the RF-SMBO algorithm, to demonstrate the framework's ability to rapidly converge on the optimal growth window.

\subsection{Feature Selection and Data Representation}
\begin{figure*}[t]
\centering
\includegraphics[width=1\textwidth]{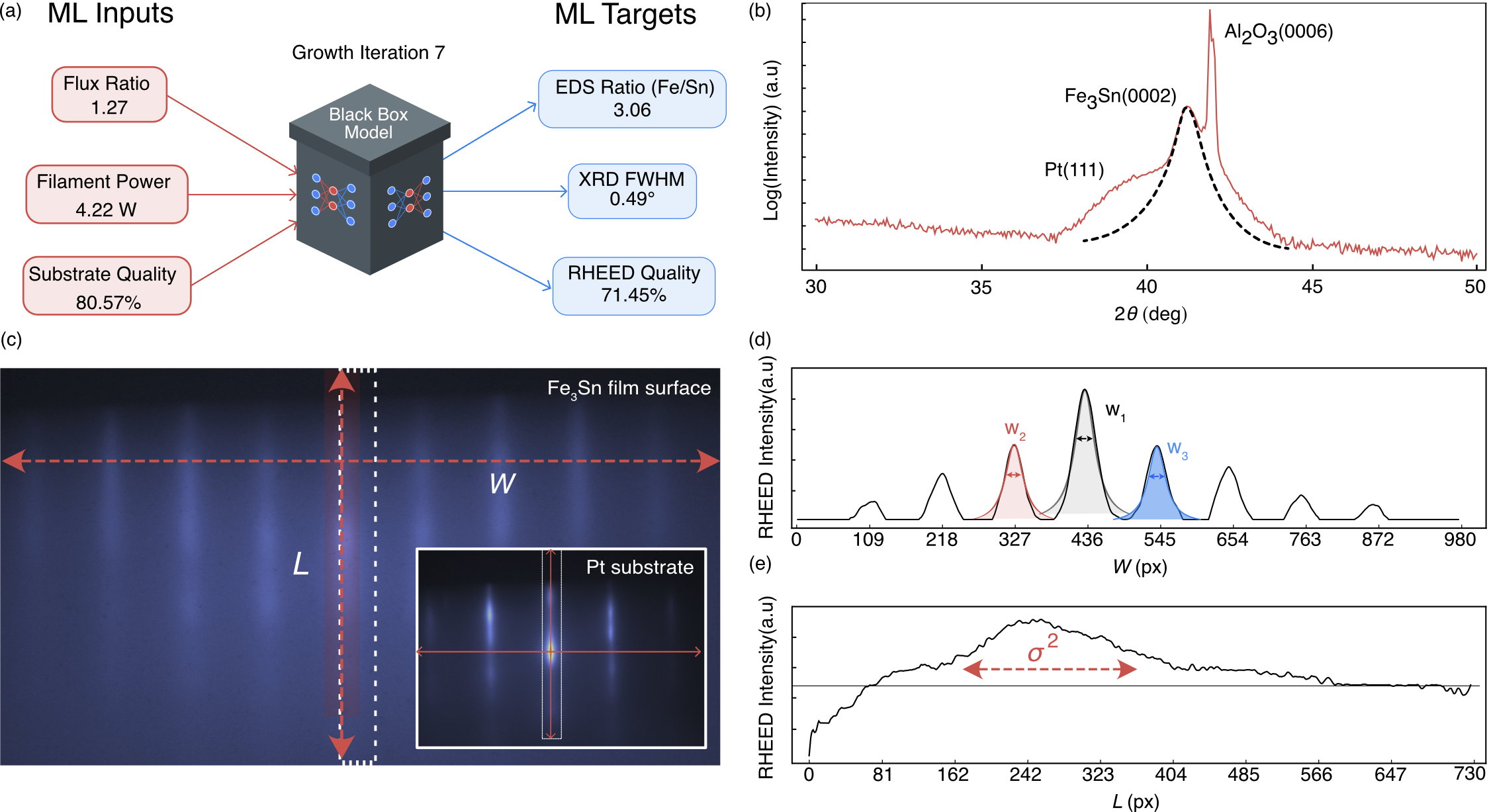}
\caption{\justifying\textbf{Feature extraction from Fe\(_3\)Sn growth.} (a) Schematic representation of the active learning "black box" surrogate model mapping independent synthesis parameters (Inputs) to structural characterization metrics (Outputs). For a representative iteration (Growth Iteration 7), parameters such as the Fe/Sn flux ratio, filament power, and initial substrate quality are fed into the model to predict and optimize the resulting holistic film quality, composed of EDS stoichiometry, XRD crystallinity, and RHEED surface morphology scores. (b) High-resolution X-ray Diffraction (XRD) $\theta-2\theta$ spectrum (solid red line) plotted on a logarithmic scale. The epitaxial $\text{Fe}_3\text{Sn}(0002)$ film peak is situated adjacent to the $\text{Al}_2\text{O}_3(0006)$ substrate peak, from which the full-width at half-maximum (FWHM) is extracted via a least-square fit, schematically shown as a black dashed line, to quantify out-of-plane crystallinity. The Pt(111) of the buffer layer is indicated. (c) {\em In-situ} Reflection High-Energy Electron Diffraction (RHEED) pattern of the $\text{Fe}_3\text{Sn}(111)$ surface after the growth is completed. The dashed orthogonal axes denote the extraction pathways for quantitative 1D intensity profiles: vertical ($L$) and horizontal ($W$). The inset shows the RHEED pattern of the underlying $\text{Pt}(111)$ substrate layer after completion of the buffer layer deposition. (d) The transversal cut of the RHEED intensity along the horizontal axis ($W$) reveals the distinct diffraction peaks, where the extracted peak widths ($w_1$-$W_3$) are utilized to quantify in-plane crystalline coherence. (e) The longitudinal cut of the RHEED intensity along the vertical axis ($L$) captures the intensity variation and spread ($\sigma^2$) of the central streak, serving as a proxy for surface roughness.}
\label{fig:fig2}
\end{figure*}

To implement the AL-MBE framework, we first construct a feature space that correlates the experimental growth parameters (ML inputs) with the quantifiable crystallographic and chemical properties of the as-grown thin films (ML targets) [see Fig.~\ref{fig:fig2}(a)]. The input feature space is defined by the primary experimental control parameters dictating the growth kinetics: the Fe-to-Sn flux ratio and the substrate heater power. We deliberately utilize the substrate heater power rather than the measured substrate temperature to eliminate calibration uncertainties inherent to infrared pyrometry. Additionally, the pre-growth crystallographic quality of the Pt(111) buffer layer, extracted via {\em in-situ} RHEED, is included as an input feature to account for substrate-induced variations in the film growth. 

To quantify the growth outcome, we extract target labels from three complementary characterization techniques: energy-dispersive X-ray spectroscopy (EDS) for film stoichiometry, X-ray diffraction (XRD) for bulk crystallinity, and RHEED for surface morphology. Because human interpretation of diffraction data is inherently subjective, we developed a suite of automated algorithms to extract physically meaningful, scalar metrics from the XRD and RHEED datasets suitable for small-data machine learning tasks\cite{chen2025semi, kaspar2025machine, yu2025multimodal}. For bulk crystallinity (XRD), we extract the full width at half maximum (FWHM) of the primary Fe$_3$Sn(0002) Bragg reflection at $2\theta \approx 41.3^\circ$, as seen in Fig.~\ref{fig:fig2}(b). This line-width is inversely proportional to the out-of-plane crystalline domain size and serves as a direct metric for structural coherence.

For surface morphology, our algorithm extracts three critical metrics that directly map to physical growth phenomena by analyzing RHEED images that were recorded after the film growth was completed. A typical RHEED image recorded on Fe$_3$Sn, such as shown in Fig.~\ref{fig:fig2}(c), is characterized by a set of vertical diffraction streaks whose characteristics allow conclusions on the crystallinity and surface morphology of the thin film. We have carried out the following analysis of the RHEED images of the as-grown Pt buffer layer and the as-grown Fe$_3$Sn film to extract information on the film crystallinity and morphology.

\vspace{0.25cm}
{\em In-plane crystallinity (Horizontal FWHM):} By fitting the primary and secondary vertical diffraction streaks along the horizontal axis $W$ to Lorentzian functions, as shown in Fig.~\ref{fig:fig2}(d), we extract the horizontal FWHM, $w_{1,2,3}$. A broadening of these streaks indicates a reduction in the in-plane crystalline domain size and the onset of structural disorder. 

\vspace{0.25cm}
{\em Film morphology (Vertical Intensity Variance):} We calculate the statistical variance of the intensity profile, seen in Fig.~\ref{fig:fig2}(e), along the vertical streak axis $L$. This physically differentiates the surface morphology: smooth, two-dimensional (2D) layer-by-layer growth yields continuous streaks (low variance), whereas three-dimensional (3D) islanding produces localized transmission diffraction spots (high variance).

\vspace{0.25cm}
{\em Structural Mosaicity (Vertical Streak Width):} The mean spatial width $\sigma^2$ of the central streak along the vertical axis $L$, as shown in Fig.~\ref{fig:fig2}(e), is extracted to quantify overall surface mosaicity.

\vspace{0.25cm}
{\em Note: Full details of the RHEED image processing and XRD analysis, including background subtraction and peak-fitting algorithms, can be accessed via the Data and Code Availabilty Section}.

\subsection{Numerical Methods: Holistic Additive Scoring}

To provide a unified scalar target for the active learning objective function, we developed a Holistic Additive Scoring framework. This maps the multi-modal characterization data onto a continuous $0$--$100$ scale, defined by a weighted sum of the structural and chemical sub-scores:
\begin{equation}
    S_{\text{Total}} = 0.50 \cdot S_{\text{RHEED}} + 0.25 \cdot S_{\text{EDS}} + 0.25 \cdot S_{\text{XRD}}
\end{equation}
The RHEED morphology score ($S_{\text{RHEED}}$) carries the highest weight ($50\%$) because the preservation of a continuous 2D film is the most stringent requirement in the MBE of Fe$_3$Sn. It is calculated via a penalty function ($P_{\text{RHEED}}$) that aggregates the extracted surface metrics:
\begin{equation}
\label{eq:penalty_RHEED}
    \begin{split}
        P_{\text{RHEED}} &=0.40 \cdot \text{FWHM}_{\text{side}} + 0.20 \cdot \text{Width}_{\text{mean}}\\
        & +0.20 \cdot \text{FWHM}_{\text{center}} + 0.20 \cdot V_{\text{curve}},
    \end{split}
\end{equation}
where $V_{\text{curve}}$ represents the vertical intensity variance. The side-streak FWHM is heavily weighted to aggressively penalize the onset of 3D roughening. The final RHEED score is defined as $S_{\text{RHEED}} = (1 - P_{\text{RHEED}}) \cdot 100$.

The chemical and bulk structural scores utilize Gaussian forgiveness curves to naturally bound the target space. The EDS stoichiometry score ($S_{\text{EDS}}$) is centered at the ideal Fe:Sn ratio of $3.0$:
\begin{equation}
    S_{\text{EDS}} = 100 \cdot \exp\left[-\frac{(r - 3.0)^2}{2\sigma_{\text{EDS}}^2}\right]
\end{equation}
where $r$ is the measured Fe:Sn ratio. Setting $\sigma_{\text{EDS}} = 0.5$ accounts for the intrinsic instrumental uncertainty of EDS measurements, preventing the AL loop from severely penalizing minor compositional fluctuations while still suppressing parasitic phases. Similarly, the XRD quality score ($S_{\text{XRD}}$) utilizes a Gaussian decay ($\sigma_{\text{XRD}} = 1.5^\circ$) applied to the Bragg peak FWHM, directly rewarding the growth of large, well-defined crystalline domains.

\subsection{Active Learning via Sequential Model-Based Optimization}

To efficiently navigate the highly constrained thermodynamic phase space of Fe$_3$Sn, we implement the active learning (AL) workflow utilizing a SMBO framework \cite{jones1998efficient, hutter2011sequential}, as schematically shown in Fig.~\ref{fig:fig3}(a). SMBO is specifically designed for the global optimization of expensive black-box functions by iteratively updating a surrogate regression model and optimizing an acquisition function to propose the next experiment. 

The AL cycle is initialized using a sparse dataset of historical growth parameters alongside the computationally extracted Substrate Master Score [calculated via $S_{\rm RHEED}$], which provides a quantitative baseline of the initial template quality. The SMBO workflow then proceeds through three iterative stages: (i) training a surrogate model on the available dataset to approximate the growth landscape, (ii) maximizing an acquisition function to identify the next optimal experimental condition, and (iii) executing the proposed growth in the physical MBE chamber, scoring the resulting film, and appending the new data point to retrain the model.

\subsubsection{Surrogate Model Selection: Overcoming Smoothness Constraints}

Given the sparsity of our experimental dataset ($N \leq 20$), choosing a suitable machine-learning surrogate is critical. Standard BO pipelines typically rely on Gaussian Process Regression (GPR) \cite{rasmussen2003gaussian}. However, as discussed previously, GPR encodes strong smoothness assumptions through its continuous covariance kernels. Consequently, it is poorly suited to model the abrupt crystallographic phase boundaries and narrow growth windows separating competing secondary phases inherent to complex binary compounds, such as Fe$_3$Sn.

To overcome this limitation, we evaluated alternative regression architectures using Leave-One-Out Cross-Validation (LOOCV) \cite{hastie2009elements} to rigorously estimate generalization error in the small-data regime. Based on the lowest LOOCV mean absolute error, we selected an RF model\cite{breiman2001random} surrogate model for the AL loop. RF is an ensemble learning method that aggregates predictions from multiple decision trees. Because it relies on hierarchical space-partitioning rather than continuous distance metrics, RF naturally captures highly non-linear relationships, sharp phase transitions, and discontinuities without assuming a smooth underlying topology. Furthermore, the bootstrap aggregation (bagging) mechanism inherently reduces variance and mitigates overfitting, making RF highly robust for the sparse, noisy experimental datasets typically found in thin film growth.

\subsubsection{Expected Improvement Acquisition}

Within the SMBO framework, the RF surrogate predicts the expected film score and also provides a measure of predictive uncertainty. Specifically, for any unexplored parameter vector $\mathbf{x}$, the predictive mean $\mu(\mathbf{x})$ is given by the average prediction of the individual decision trees, while the predictive standard deviation $\sigma(\mathbf{x})$ is derived from the variance across the trees in the forest.

To determine the next best experimental condition for growing a thin film with improved material properties, these statistical outputs are fed into the Expected Improvement (EI) acquisition function \cite{jones1998efficient}. Instead of blindly exploring the parameter space, EI calculates the statistical probability that a specific set of parameters will yield a score higher than the current historical maximum, $f^*$:
\begin{equation}
    \text{EI}(\mathbf{x}) = \mathbb{E}\left[\max(f(\mathbf{x}) - f^* , 0)\right]
\end{equation}
Assuming a roughly Gaussian distribution of the tree predictions at $\mathbf{x}$, the EI can be computed analytically:
\begin{equation}
    \text{EI}(\mathbf{x}) = (\mu(\mathbf{x}) - f^*)\Phi(Z) + \sigma(\mathbf{x})\phi(Z)
\end{equation}
where $Z = (\mu(\mathbf{x}) - f^*)/\sigma(\mathbf{x})$, and $\Phi$ and $\phi$ are the cumulative distribution function and probability density function of the standard normal distribution, respectively. 

The EI function thus naturally resolves the exploration-exploitation dilemma: the first term drives exploitation by favoring regions where the predicted mean $\mu(\mathbf{x})$ is high, while the second term drives exploration by favoring regions with high uncertainty $\sigma(\mathbf{x})$. By maximizing $\text{EI}(\mathbf{x})$ across the phase space, the SMBO algorithm dynamically identifies the most informative parameters for the next MBE growth, efficiently driving the system toward the optimal growth condition [see Fig.~\ref{fig:fig1}(a)] while minimizing human intervention and expense of experimental resources.

\section{Results and Discussion}

\subsection{Surrogate Model Performance: The Illusion of Smoothness}

To empirically validate our choice of surrogate model within the active learning loop, we compared the predictive landscapes generated by RF against GPR, the standard model of conventional Bayesian optimization previously used for Al-MBE~\cite{ohkubo2021realization, zheng2025self}. Evaluated via LOOCV on the sparse 17-iteration dataset, GPR and RF yielded seemingly comparable Mean Absolute Errors MAE $\approx 9.92$ and $10.16$, respectively. However, an analysis of their predicted landscapes reveals that GPR's marginally lower error is a mathematical artifact of kernel collapse [see Fig.~\ref{fig:fig3}(b)]. Because epitaxial film formation stability is bounded by steep thermodynamic drop-offs, the continuous radial basis function kernel of the GPR smoothed over these abrupt boundaries, flattening its predictions toward the global dataset mean.

In contrast, the RF ensemble accurately mapped the underlying thermodynamic topology [blue line in Fig.~\ref{fig:fig3}(b)]. Rather than falling into the mean-reversion trap of smooth continuous models, its tree-based architecture successfully captured the non-linear boundaries of the growth landscape without artificially suppressing variance. This empirical result suggests that pure BO approaches, previously studied for active learning MBE of single element materials\cite{ohkubo2021realization, zheng2025self}, are ill-equipped for complex binary phase spaces, whereas the RF framework provides the reliable topology and uncertainty quantification ($\sigma$) required to successfully drive the AL loop.

\subsection{The Acquisition Landscape and Optimal Growth Conditions}

Using the validated RF surrogate, we evaluated the EI acquisition function across the unexplored parameter space. Figure~\ref{fig:fig3}(c) presents a subspace of the resulting acquisition landscape spanning the Fe/Sn flux ratio and substrate filament power. Rather than a singular, sharp global maximum, the landscape reveals a "Plateau of Success"--a well-defined, robust thermodynamic window centered around a filament power of approx~$3.6$\,W and a Fe:Sn atomic flux ratio of approx.~$0.8$ [indicated by black dashed box].

The existence of this plateau, which spans a temperature range of approx. 30\,K is physically significant. It shows that within this specific optimal window, the crystalline quality of Fe$_3$Sn thin films is robust to minor fluctuations in deposition parameters--a critical requirement for practical thin film synthesis. As indicated by the historical growth coordinates [white circles in Fig.~\ref{fig:fig3}(c)], human involvement, based on the analysis of crystallographic and stoichiometric film data, successfully guided the growth conditions toward the Plateau of Success using 17 growth iterations only. Below, we will show that our active learning loop will autonomously guide subsequent experiments onto this plateau, with the global maximum of the acquisition landscape (yellow star) continuously refining the proposed target.

\subsection{Physical Insights from Feature Importance}

\begin{figure*}[t]
\centering
\includegraphics[width=\textwidth]{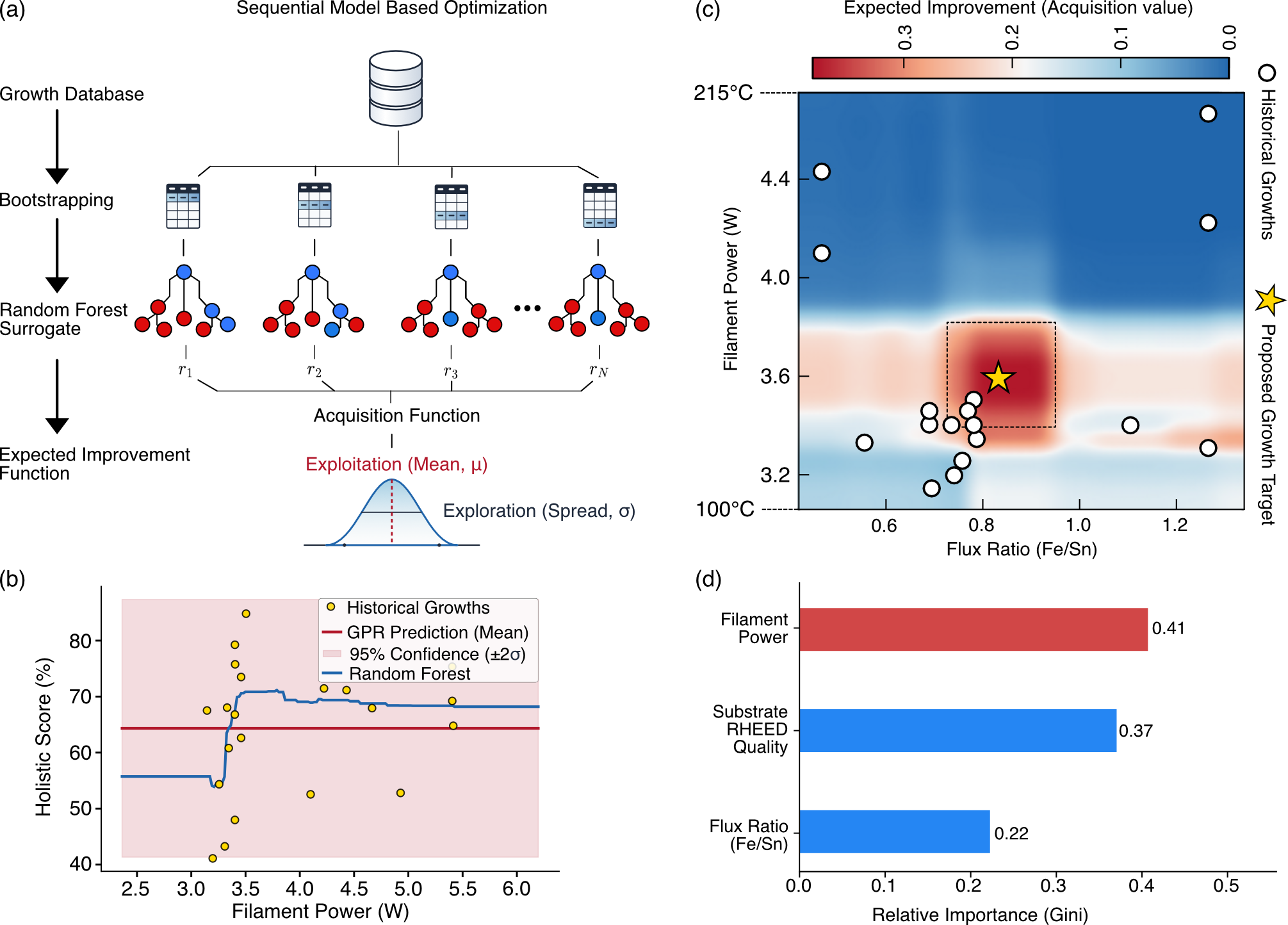}
\caption{\justifying{\bf Random Forest based SMBO framework, performance, and feature importance:} (a) Schematic representation of the SMBO framework. Historical growth data from the database is sampled with replacement (bootstrapping) to train an ensemble of independent decision trees of a Random Forest model. The ensemble average ($\mu$) provides the prediction (exploitation), while the variance ($\sigma$) across trees provides the uncertainty (exploration), which together formulate the acquisition function for the expected improvement analysis. (b) 1D cross-section evaluating model predictions of the Holistic Score as a function of Filament Power. The Gaussian Process Regressor (GPR) exhibits kernel collapse on the sparse dataset, defaulting to the global mean (red line) with large uncertainty (pink shaded region representing $\pm2\sigma$ ). Conversely, the Random Forest (blue line) robustly models the non-linear parameter space. (c) The Expected Improvement acquisition landscape evaluated across the 2D parameter space of Filament Power and Flux Ratio. Historical growths are marked with white circles, and the global maximum—representing the optimal proposed target for the next active learning iteration—is indicated by a gold star. The "Plateau of Success" as the area of optimum growth condition is highlighted by a black dashed box. (d) Relative feature influence (Gini impurity \cite{altmann2010permutation}) extracted from the Random Forest model.}
\label{fig:fig3}
\end{figure*}

Beyond serving as an optimization engine, the RF surrogate extracts physical insights into the growth parameters, that is, ML inputs. Figure~\ref{fig:fig3}(d) quantifies the relative influence of the three input parameters on the holistic film quality using Gini impurity reduction \cite{altmann2010permutation}. Substrate filament power (Importance $\approx 0.41$) and the initial substrate RHEED quality ($0.37$) dominate over the Fe/Sn flux ratio, which exerts a comparatively minor influence ($0.22$).

This hierarchy mirrors the physical realities of the growth dynamics. Filament power—a direct proxy for substrate temperature—determines adatom surface mobility and governs the competition between kinetics and thermodynamics. The surrogate model autonomously identified temperature as the most critical bottleneck: temperatures that are too low result in kinetically limited growth, such as islanding, and mixed phase formation with Fe$_3$Sn$_2$~\cite{shen2022thermodynamical}, while temperatures that are too high cause volatile Sn desorption and parasitic phase formation as well as alloying with the Pt buffer layer~\cite{cheng2022atomic}.

Equally Interesting is the high importance of the initial substrate quality, which suggests that no amount of post-deposition parameter tuning can compensate for a poor starting surface. If the surface morphology of the initial template deviates from a flat terraced surface, as quantified through continuous streaks in the RHEED image, the epitaxial film cannot achieve high crystalline quality. This finding directs future experimental design, suggesting that improving substrate preparation protocols yields significantly higher returns than exhaustive fine-tuning of elemental fluxes. 

Finally, the relatively low importance of the flux ratio indicates that, provided the temperature is precisely tuned to the optimal growth window identified in Fig.~\ref{fig:fig3}(c), the crystalline and stoichiometric properties of the Fe$_3$Sn film are robust against minor stoichiometric fluctuations in the atomic beam fluxes, contrasting with the metastable line-compound nature of Fe$_3$Sn in bulk synthesis~\cite{shen2022thermodynamical}, where deviations from ideal stoichiometry could result in mixed phase compounds.

\subsection{Dynamic Convergence of the Active Learning Loop}

\begin{figure*}[t]
\centering
\includegraphics[width=1.0\textwidth]{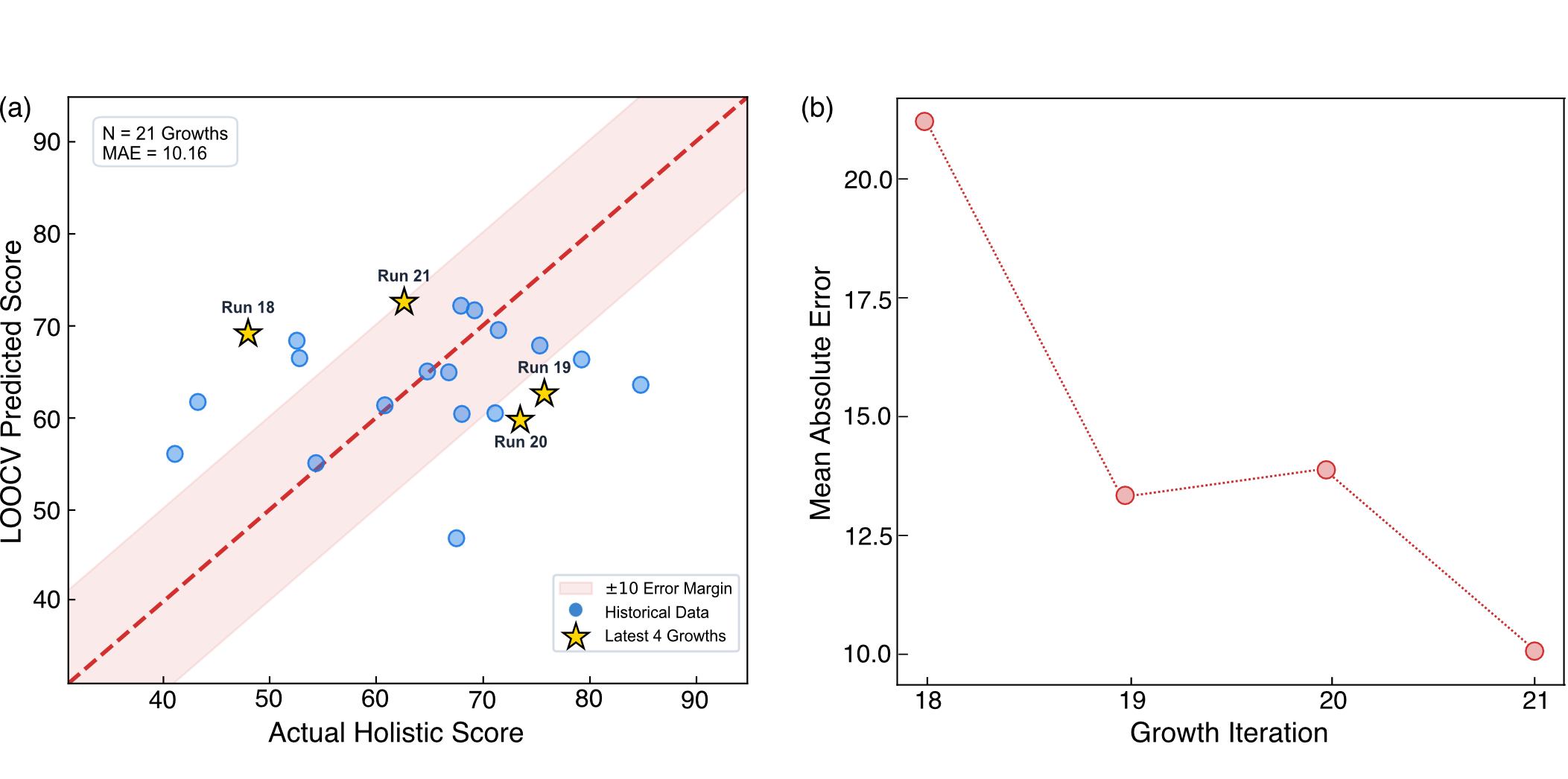}
\caption{\justifying\textbf{Implementing active learning MBE of complex quantum materials:} (a) The main parity plot compares the leave-one-out cross-validation (LOOCV) predicted holistic scores against the actual experimental scores for $N=21$ total growths. The dashed red diagonal represents the ideal prediction ($y=x$), bounded by a shaded $\pm10$ error margin. Historical data points are shown as blue circles, while the latest four active learning iterations (Runs 18–21) are highlighted as gold stars. (b) Shown is the MAE as a function of the growth number for the final four iterations.}
\label{fig:fig4}
\end{figure*}

The overarching goal of the SMBO framework is to minimize experimental iterations while maximizing predictive accuracy. The predictive performance of the RF surrogate is visualized in the parity plot [Fig.~\ref{fig:fig4}(a)], which compares the LOOCV-predicted scores against actual experimental scores. The data tightly clusters along the ideal 1:1 diagonal, with a global MAE of $10.16$ on a $100$-point scale. This precision is sufficient to reliably distinguish between optimal, moderate, and parasitic growth regimes under extreme data scarcity.

To establish AL MBE of complex quantum materials shown in Fig.~\ref{fig:fig1}(b), we have used our SMBO model framework to predict the next best growth parameters, performed the MBE of Fe$_3$Sn using the updated recipe, and characterized the resulting thin films. The resulting film properties were fed back into the surrogate model retraining and a total of four growth iterations (18--21) were performed. The dynamic learning capability of the SMBO loop is most explicitly demonstrated by tracking the absolute prediction error during these final active learning iterations, shown in Fig.~\ref{fig:fig4}(b). Initially, Run 18 exhibits a sharp spike in error, peaking above $20.0$. Rather than a model failure, this reflects the Expected Improvement (EI) acquisition function correctly managing the exploration-exploitation tradeoff. The algorithm deliberately probed a high-uncertainty region of the thermodynamic phase space to map uncharted growth boundaries. 

Crucially, as the results of Run 18 and 19 were fed back into the RF surrogate, the model rapidly assimilated this new physical information. The prediction error plummeted to $13.2$ in Run 19, stabilized through Run 20, and dropped to a highly accurate $9.9$ by Run 21. This rapid error reduction—halved within only four iterations—proves that the RF surrogate model successfully prevents premature convergence to local optima while rapidly zeroing in on the "Plateau of Success."

\section{Discussion and Conclusion}

The performance of this RF-driven active learning framework sets a new benchmark for data-efficient optimization of complex materials within closed-loop synthesis. While existing proof-of-principle demonstrations of closed-loop AL in MBE have successfully removed human intuition from the growth cycle, they have predominantly been restricted to structurally simple, single-element or pseudo-elemental thin films (e.g., Ag or TiN)~\cite{ohkubo2021realization, zheng2025self}. These systems are characterized by forgiving, low-dimensional optimization landscapes. Conversely, when standard GPR-based Bayesian optimization has been applied outside an active learning framework to more complex functional materials, such as perovskite oxides (e.g., SrRuO$_3$ and SrTiO$_3$), convergence typically demanded $24$ to $44$ experimental iterations~\cite{wakabayashi2019machine, wakabayashi2022bayesian, wakabayashi2023stoichiometric}. These findings contrast with our SMBO approach which achieves robust, closed-loop optimization of the topological Weyl ferromagnet Fe$_3$Sn in only $21$ total iterations. This rapid convergence validates our methodological departure from standard BO toward an RF-based SMBO for AL MBE.

In conclusion, we have demonstrated a highly data-efficient active learning framework capable of navigating the constrained, multi-dimensional thermodynamic phase spaces of complex quantum materials. By replacing the continuous surrogate models of standard Bayesian optimization with a Random Forest architecture, our SMBO approach successfully navigated the complex optimization landscape of the metastable topological Weyl ferromagnet Fe$_3$Sn. Guided by an Expected Improvement acquisition, the closed-loop system autonomously identified a robust "Plateau of Success" as the optimum growth window and halved the predictive error within only four additional active learning iterations. Furthermore, the random forest surrogate model extracted critical physical insights without human bias, identifying substrate temperature and quality as the dominant thermodynamic gatekeepers of epitaxial film growth. 

Looking forward, we anticipate that the inclusion of further input parameters, such as deposition rate, as well as target parameters, such as the analysis of time-dependent RHEED spectra recorded during thin film growth and a more detailed crystallographic characterization [$\omega$- and $\phi$-scan] of the Fe$_3$Sn thin films, will likely result in an even further reduction of the model error. It will also be interesting to optimize the holistic additive scoring of the output data, which has been left for future studies. More broadly, the data efficiency and algorithmic robustness of the active learning framework presented in this study remove a critical computational barrier toward the realization of fully autonomous thin film materials synthesis. While our current protocol successfully automates the optimization and decision-making process, the natural next step is the physical integration of this SMBO framework with automated effusion cell control, robotic sample handling, and real-time \textit{in-situ} characterization. Equipping such "self-driving" MBE laboratories with active learning frameworks will facilitate autonomous materials discovery, expediting the exploration and thin film growth of complex quantum materials.

\section*{Data and Code Availability}

To ensure the reproducibility of this project, all data and custom Python algorithms developed for this workflow have been open-sourced. The complete suite of tools, including the RHEED digital signal processing scripts, active learning models, and XRD raw data with OriginPro files, can be accessed via the following unified GitHub repository: \url{https://github.com/raghu0415/Fe3SnMLfolder.git}

\section*{Acknowledgments}
We acknowledge valuable discussions with Shiming Lei. This work was supported by the Hong Kong Research Grants Council (Grant Nos.\,26304221, 16302422, 16302624, 16304525, and C6033-22G) and the Croucher Foundation (Grant No.\,CIA22SC02).

\bibliography{bibliography}

\end{document}